\documentclass[preprint,aps,prd,nofootinbib]{revtex4-1}
\usepackage[english]{babel}
\usepackage{verbatim}
\usepackage{mathtools}
\usepackage{amssymb}
\usepackage{amsthm}
\usepackage{amsmath}

\newcommand{\mc}{\mathcal}
\newcommand{\MHV}{\mathrm{MHV}}
\newcommand{\NMHV}{\mathrm{NMHV}}
\newcommand{\NKMHV}{\mathrm{N}^K\mathrm{MHV}}
\newcommand{\gMHV}{\overline{\mathrm{MHV}}}
\newcommand{\avg}[1]{\langle#1\rangle}
\newcommand{\Avg}[1]{[#1]}

\newcommand{\Tr}{\operatorname{Tr}}
\newcommand{\hc}{\mathrm{h.c.}}

\renewcommand{\P}{\mc{P}}
\newcommand{\oP}{\hat{\mc{P}}}
\newcommand{\oQ}{\hat{\mc{Q}}}
\newcommand{\A}{\mc{A}}

\newcommand{\bra}[1]{\langle #1 |}

\newcommand{\expval}[3]{\langle #1|#2|#3 \rangle}

\begin{document}
\date{\today}
\author{Mads S{\o}gaard}
\title{Bilocal phase operators in $\beta$-deformed super Yang-Mills}
\affiliation{Niels Bohr International Academy and Discovery Center,
Niels Bohr Institute, Blegdamsvej 17, DK-2100 Copenhagen, Denmark}

\begin{abstract}
We present tree-level scattering amplitudes in $\beta$-deformed super Yang-Mills
theory in terms of new generating functions, derived by construction of a phase
operator and application thereof to the $\mc N = 4$ superamplitudes. The
technique is explicitly illustrated for the MHV and NMHV sectors. Along these
lines we propose a phase representation of the $\mc N = 4$ superconformal
algebra realized on deformed amplitudes in the planar limit. Validity of the MHV
vertex expansion is proven and a connection to non-planar multi-loop unitarity
cuts is established. Our derivations are also compatible with the related
$\gamma$-deformation.
\end{abstract}

\maketitle

%%%%%
\section{Introduction}
Maximally supersymmetric ($\mc N = 4$) super Yang-Mills theory in four
dimensions is a very special quantum field theory in several ways. First of all,
it has surprisingly simple and well-behaved amplitudes. The $\mc N = 4$
supermultiplet is CPT self-conjugate and in addition the theory is both
classically and quantum mechanically conformally invariant because the
renormalization group $\beta$-function vanishes identically. Even more
remarkably, although gravitational interactions are absent, the $\mc N = 4$
theory is intimately related to a supergravity theory via the celebrated AdS/CFT
correspondence.

Multiloop $\mc N = 4$ scattering amplitudes are very elegantly studied within
the on-shell superspace formalism \cite{Bern:2009xq,Sogaard:2011pr,
Bianchi:2008pu,Elvang:2008na,Georgiou:2004by,Elvang:2008vz,Elvang:2011fx,
ArkaniHamed:2008gz,Drummond:2008vq,Brandhuber:2008pf,Drummond:2008cr,
Drummond:2009fd,Drummond:2008bq}. The principle is to arrange the entire
supermultiplet as a superfield expanded in Grassmann variables. All possible
scattering combinations are realized by formation of superamplitudes, defined as
generating functions of $n$ copies of superfields corresponding to the external
legs. Single amplitudes are then projected out by unique strings of Grassmann
differential operators or pieced together in a graphical framework.
Superamplitudes for general particle and helicity configurations may be
constructed with the Britto-Cachazo-Feng-Witten (BCFW) on-shell recursion
relations
\cite{Britto:2004ap,Britto:2005fq,ArkaniHamed:2008gz,Brandhuber:2008pf}.

It is of course very interesting to examine features of the $\mc N = 4$ theory
in more general settings. Super Yang-Mills theories with less than maximal
supersymmetry do not possess all of the amazing properties mentioned, although
the superspace formalism can be generalized
\cite{Elvang:2011fx,Lal:2009gn,Sogaard:2011pr}. However, exactly marginal
deformations of the $\mc N = 4$ super Yang-Mills theory preserving only reduced
amount of supersymmetry in particular inherit conformal invariance
\cite{Leigh:1995ep}, and have recently attracted considerable attention
especially catalyzed by the AdS/CFT duality. Exactly marginal deformations have
also been subject to intense investigations within the perturbative regime
\cite{Khoze:2005nd,Mansson:2008xv,Lunin:2005jy,Ananth:2006ac,Ananth:2007px,
Elmetti:2006gr,Elmetti:2007up,Kazakov:2007dy,Ihry:2008gm,Bork:2007bj,Oz:2007qr,
Mauri:2005pa,Fiamberti:2010fw,Madhu:2007ew,Kulaxizi:2006zc,Bundzik:2006jz}. In
particular, it has been of special interest to understand how scattering
amplitudes are modified by the so-called $\beta$-deformation. It can be shown
that conformal invariance of the planar $\beta$-deformed super Yang-Mills theory
requires reality of $\beta$, the deformation parameter. The theory enjoys only
$\mc N = 1$ supersymmetry and a global $U(1)_1\times U(1)_2$ flavor symmetry,
and may be formulated with both symmetries manifest in the $\mc N = 1$
superspace formalism in terms of three charged chiral superfields and neutral
vector superfield, and a star product operation which incorporates the charges
of the fields under the flavor symmetry. Using this Lagrangian approach the
$\beta$-deformed Feynman rules of the deformed theory may be obtained, and the
behaviour of its scattering amplitudes elucidated \cite{Khoze:2005nd}. It turns
out that the $\mc N = 4$ and $\beta$-deformed theories are very similar.

Motivated by the developments in exactly marginal deformations and inspired by
how powerfully on-shell scattering amplitudes are constructed from analyticity
and unitarity in the $\mc N = 4$ on-shell framework, we recast the deformed
theory in terms of generating functions.

%%%%%
\section{Notation and conventions}
Several modern techniques apply to present gauge theory scattering amplitudes 
very effectively. With all states in the adjoint representation of the gauge
group, say $SU(N_c)$, any tree-level amplitude can be color decomposed as
\begin{align}
{\mathbb A}^{\textrm{tree}}_n(1,2,\dots,n) = g^{n-2}\sum_{\sigma\in S_n/Z_n}
\Tr\left(T^{a_{\sigma(1)}}T^{a_{\sigma(2)}}\cdots T^{a_{\sigma(n)}}\right)
A_n^{\textrm{tree}}(\sigma(1),\sigma(2),\dots,\sigma(n))\;,
\end{align}
where $g$ is the gauge coupling constant and
$A_n^{\textrm{tree}}(\sigma(1),\sigma(2),\dots,\sigma(n))$ are partial
amplitudes soaking up the entire kinematical structure corresponding to a
particular ordering of the $n$ external legs encoded by the trace of the gauge
group generators $T^a$. The sum is over all permutations $S_n$ with trace
preserving cyclic permutations $Z_n$ modded out.

The partial amplitudes are written in terms of Lorentz invariant, little group
covariant holomorphic and antiholomorphic spinor products. In order to
distinguish the two chiralities we use angle and square brackets, and define
\begin{align}
\avg{ij} = -\avg{ji} \equiv 
\epsilon_{\alpha\beta}\lambda_i^\alpha\lambda_j^\beta\;, \quad
\Avg{ij} = -\Avg{ji} \equiv 
\epsilon_{\dot\alpha\dot\beta}
\tilde\lambda_i^{\dot\alpha}\tilde\lambda_j^{\dot\beta}
\end{align}
for commuting spinors $\lambda_i^\alpha$ and $\tilde\lambda_i^{\dot\alpha}$ 
related to momentum by 
$p_i^{\alpha\dot\alpha} = \lambda_i^\alpha\lambda_i^{\dot\alpha}$.

%%%%%
\section{On-shell $\mc N = 4$ supersymmetry}
The $\mc N = 4$ gauge multiplet has the unique property of CPT self-conjugacy
which implies that all on-shell states can be assembled by a single holomorphic
superfield $\Phi(p,\eta)$. It is expanded in Grassmann variables $\eta_a$ where
$a = 1,\dots,4$ is a fundamental index of $SU(4)_R$, the $R$-symmetry group of
the theory. Within this setup the sixteen physical states transform in $r$-rank
antisymmetric tensor representations as two gluons $g_+$ and $g^{abcd}_-$, four
fermion pairs $f^a_+$ and $f^{abc}_-$, plus six real scalars $s^{ab}$. The 
tensor rank $r$ and particle helicity $h$ are related through $2h = 2-r$.

The superfield is \cite{Drummond:2008cr} (repeated indices are summed)
\begin{align}
\Phi(p,\eta) = g_++\eta_af^a_++\frac{1}{2!}\eta_a\eta_bs^{ab}
               +\frac{1}{3!}\eta_a\eta_b\eta_c f^{abc}_-
               +\frac{1}{4!}\eta_a\eta_b\eta_c\eta_dg^{abcd}_-\;.
\end{align}
There also exists an antiholomorphic superfield
\begin{align}
\label{N=4ANTISUPERFIELD}
\tilde\Phi(p,\tilde\eta) = 
g^-+\tilde\eta^af_a^-+\frac{1}{2!}\tilde\eta^a\tilde\eta^bs_{ab}
+\frac{1}{3!}\tilde\eta^a\tilde\eta^b \tilde\eta^c f^+_{abc}
+\frac{1}{4!}\tilde\eta^a\tilde\eta^b \tilde\eta^c\tilde\eta^dg^+_{abcd}\;,
\end{align}
linked to $\Phi(p,\eta)$ by the Grassmann Fourier transform, but with the exact 
same particle content encoded. Due to this equivalence either representation may 
be preferred. In this paper we reserve the holomorphic and antiholomorphic 
descriptions for $\MHV$ and $\gMHV$ amplitudes respectively.

\subsection{MHV superamplitudes}
Proliferation of scattering amplitudes in $\mc N = 4$ super Yang-Mills theory is
handled by introduction of superamplitudes, which are functions of $n$ copies of
the superfield, one for each external leg. The full $n$-point tree-level
superamplitude is organized ascendingly according to Grassmann degree in steps
of four,
\begin{align}
\A_n(\lambda,\tilde\lambda,\eta) = \A(\Phi_1\cdots\Phi_n) = 
\A_n^{\MHV}+\A_n^{\NMHV}+\cdots+\A_n^{\gMHV}\;,
\end{align}
ranging from eight $\eta$'s to $4n-8$. It follows in particular that all MHV 
amplitudes may be packaged into a generating function, also referred to as the 
MHV superamplitude. Each term thus corresponds to a regular scattering amplitude 
involving gluons, fermions and scalars. It is convention to extract the MHV
sector from the full superamplitude as an overall factor.

The MHV tree-level superamplitude is in addition to the well-known overall 
momentum conservation proportional to an eightfold Grassmann delta function 
conserving total supermomentum 
$Q^\alpha_a\equiv\sum_{j=1}^n\lambda_j^\alpha\eta_{ja}$, and is defined by 
\cite{Bern:2009xq}
\begin{align}
\A^{\mathrm{MHV}}_n = 
i\frac{(2\pi)^4\delta^{(4)}(\sum_{i=1}^n p_i)}{\prod_{r=1}^{n}\avg{r(r+1)}}
\delta^{(8)}\bigg(\sum_{j=1}^n\lambda_j^\alpha\eta_{ja}\bigg)\;.
\end{align}
For calculational purposes it proves advantageous to expand the Grassmann delta
function present in the MHV superamplitude as a sum of monomials in the
$\eta$'s, first in all possible values of the group index,
\begin{align}
\delta^{(8)}\bigg(\sum_{j=1}^n\lambda_j^\alpha\eta_{ja}\bigg) =
\prod_{a=1}^4\delta^{(2)}\bigg(\sum_{j=1}^n\lambda_j^\alpha\eta_{ja}\bigg)\;,
\end{align}
and then using $\delta(\eta) = \eta$ for Grassmann variables,
\begin{align}
\delta^{(8)}\bigg(\sum_{j=1}^n\lambda_j^\alpha\eta_{ja}\bigg) = 
\prod_{a=1}^4\sum_{i<j}\avg{ij}\eta_{ia}\eta_{ja}\;.
\end{align}
Frequently we write holomorphic and antiholomorphic spinor products of 
supermomenta of the individual legs defined by
\begin{align}
\avg{q_{ia}q_{ja}}\equiv \eta_{ia}\avg{ij}\eta_{ja}\;, \qquad
\Avg{\tilde q_i^a\tilde q_j^a}\equiv \tilde\eta_i^a\Avg{ij}\tilde\eta_j^a\;.
\end{align}
Consequently, the MHV generating function reaches the very clean form
\begin{align}
\label{MHVSUPERAMPLITUDE}
\A^{\mathrm{MHV}}_n = 
i\frac{\prod_{a=1}^4\sum_{i<j}\avg{q_{ia}q_{ja}}}
{\prod_{r=1}^{n}\avg{r(r+1)}}\;,
\end{align}
with four-momentum conservation stripped.

For four external legs, some simple examples of MHV component amplitudes are
\begin{align}
\label{SIMPLEMHV1}
A_4^{\MHV}(
1^-_{g^{1234}},2^-_{g^{1234}},3^+_{g},4^+_{g}) &=
i\frac{\prod_{a=1}^4\avg{q_{1a}q_{2a}}}{\avg{12}\avg{23}\avg{34}\avg{41}}\;, \\
\label{SIMPLEMHV2}
A_4^{\MHV}(
1^-_{g^{abcd}},2^-_{f^{abc}},3^+_{f^d},4^+_{g}) &=
i\frac{\avg{q_{1a}q_{2a}}\avg{q_{1b}q_{2b}}\avg{q_{1c}q_{2c}}\avg{q_{1d}q_{3d}}}
{\avg{12}\avg{23}\avg{34}\avg{41}}\;, \\
A_4^{\MHV}(
\label{SIMPLEMHV3}
1^-_{f^{abc}},2^-_{f^{abd}},3_{s^{cd}},4^+_{g}) &=
i\frac{\avg{q_{1a}q_{2a}}\avg{q_{1b}q_{2b}}\avg{q_{1c}q_{3c}}\avg{q_{2d}q_{3d}}}
{\avg{12}\avg{23}\avg{34}\avg{41}}\;.
\end{align}

Analogous to the MHV superamplitude we may use the antiholomorphic superfields
\eqref{N=4ANTISUPERFIELD} to build a $\gMHV$ generating function. The $\gMHV$ 
superamplitude conserves total conjugate supermomentum
$\tilde Q^{\dot\alpha a} = 
\sum_{i=1}^n\tilde\lambda_i^{\dot\alpha}\tilde\eta_i^a$ 
and is apart from an ordinary momentum conserving delta function given by
\begin{align}
\label{BARMHVSUPERAMPLITUDE}
\A^{\mathrm{\gMHV}}_n = 
i(-1)^n\frac{\delta^{(8)}
(\sum_{i=1}^n\tilde\lambda_i^{\dot\alpha}\tilde\eta_i^a)}
{\prod_{r=1}^{n}\Avg{r(r+1)}} =
i(-1)^n \frac{\prod_{a=1}^4\sum_{i<j}\Avg{\tilde q_i^a \tilde q_j^a}}
{\prod_{r=1}^n\Avg{r(r+1)}}\;.
\end{align}
It is mapped from the $\tilde\eta$-coordinates to the untilded superspace using
the Grassmann Fourier transform realized by the $n$-leg operator
\begin{align}
\hat{\mc F}\bullet \equiv \int\prod_{i,a}d\tilde\eta_i^a\exp\left(
\sum_{b,j}\tilde\eta^b_j\eta_{jb}\right)\bullet\;.
\end{align}

As a first example of $\gMHV$ component amplitudes consider the equivalent
reinterpretation of the MHV amplitudes \eqref{SIMPLEMHV1}-\eqref{SIMPLEMHV3} in 
the tilded superspace,
\begin{align}
A_4^{\gMHV}(
1^-_g,2^-_g,3^+_{g_{1234}},4^+_{g_{1234}}) &=
i\frac{\prod_{a=1}^4\Avg{\tilde q_3^a\tilde q_4^a}}
{\Avg{12}\Avg{23}\Avg{34}\Avg{41}}\;, \\
A_4^{\gMHV}(
1^-_g,2^-_{f_{d}},3^+_{f_{abc}},4^+_{g_{abcd}}) &=
i\frac{\Avg{\tilde q_{3a}\tilde q_{4a}}\Avg{\tilde q_{3b}\tilde q_{4b}}
\Avg{\tilde q_{3c}\tilde q_{4c}}\Avg{\tilde q_{2d}\tilde q_{4d}}}
{\Avg{12}\Avg{23}\Avg{34}\Avg{41}}\;, \\
A_4^{\gMHV}(
1^-_{f_{d}},2^-_{f_{c}},3_{s_{ab}},4^+_{g_{abcd}}) &=
i\frac{\Avg{\tilde q_{3a}\tilde q_{4a}}\Avg{\tilde q_{3b}\tilde q_{4b}}
\Avg{\tilde q_{2c}\tilde q_{4c}}\Avg{\tilde q_{1d}\tilde q_{4d}}}
{\Avg{12}\Avg{23}\Avg{34}\Avg{41}}\;.
\end{align}

\subsection{The NMHV sector}
The supersymmetric BCFW recursion relations
\cite{ArkaniHamed:2008gz,Brandhuber:2008pf} generate all tree-level
superamplitudes in $\mc N = 4$ super Yang-Mills theory in terms of nested sums
of dual superconformal invariants \cite{Drummond:2008cr} from just $\MHV$ and
$\gMHV$ amplitudes. The essense is to deform the on-shell superspace by a
supershift and recover the physical amplitude by residue calculus. In this paper
it suffices to examine in detail only the $\NMHV$ superamplitude. It comes with
the very compact result \cite{Drummond:2008cr}
\begin{align}
\A_n^{\NMHV} = \A_n^{\MHV}\sum_{1<s<t<n}R_{n;st}\;,
\end{align}
where, however, $R_{n;st}$ are complicated dual superconformal invariants 
(see e.g. \cite{Drummond:2008vq}),
\begin{align}
\label{DUALSUPERCONFORMALINV}
R_{n;st} = \frac{\avg{s(s-1)}\avg{t(t-1)}\delta^{(4)}(\Xi_{n;st})}
{x_{st}^2\expval{n}{x_{ns}x_{st}}{t}\expval{n}{x_{ns}x_{st}}{t-1}
\expval{n}{x_{nt}x_{ts}}{s}\expval{n}{x_{nt}x_{ts}}{s-1}}\;,
\end{align}
depending on another intricate Grassmann valued object, $\Xi_{n;st}$, defined by
\begin{align}
\label{NMHVXI}
\Xi_{n;st} = \sum_{i=t}^{n-1}\expval{n}{x_{ns}x_{st}}{i}\eta_{ia}+
\sum_{i=s}^{n-1}\expval{n}{x_{nt}x_{ts}}{i}\eta_{ia}\;.
\end{align}

Equipped with all necessary machinery of $\mc N = 4$ superamplitudes we are now 
ready to consider applications in $\beta$-deformed super Yang-Mills theory.

%%%%%
\section{$\beta$-deformed super Yang-Mills theory}
Marginal deformations of conformally invariant supersymmetric gauge theories
were first systematically studied by Leigh and Strassler \cite{Leigh:1995ep},
and have subsequently been analyzed extensively both perturbatively and at
strong coupling in \cite{Khoze:2005nd,Mansson:2008xv,Lunin:2005jy,Mauri:2005pa,
Ananth:2006ac,Ananth:2007px,Elmetti:2006gr,Elmetti:2007up,Kazakov:2007dy,
Ihry:2008gm,Bork:2007bj,Oz:2007qr,Fiamberti:2010fw,Madhu:2007ew,Kulaxizi:2006zc,
Bundzik:2006jz} just to mention a few. 

Before exactly marginal deformations are treated, we shall first pay brief 
attention to the Lagrangian formulation of the undeformed $\mc N = 4$ super 
Yang-Mills theory. The standard $\mc N = 1$ superspace formalism applies by 
construction to field theories with $\mc N = 1$ supersymmetry. But since 
$\mc N > 1$ supersymmetric theories may be reduced to multiple unextended 
supermultiplets coupled together, the $\mc N = 1$ framework proves useful in a 
much richer class of theories. The $\mc N = 4$ particle content decomposes into 
one $\mc N = 1$ vector multiplet and three $\mc N = 1$ chiral multiplets. The
Lagrangian can be written \cite{Gates:1983nr,Fiamberti:2010fw}
\begin{align}
\label{N=4SYMLAGRANGIAN}
\mc L_{\mc N = 4} = &
\int d^2\theta d^2\bar\theta\Tr e^{-gV}\bar\Phi^ie^{gV}\Phi_i \nonumber \\ & {}
+\frac{1}{2g^2}\int d^2\theta\Tr W^\alpha W_\alpha
+\left\{g\int d^2\theta\Tr\Phi_1[\Phi_2,\Phi_3]+\hc\right\}\;,
\end{align}
where $V$ is the vector superfield and $\Phi_1$, $\Phi_2$ and $\Phi_3$ are the 
three chiral superfields. The kinetic term for the vector superfield involves as 
usual the superfield strength defined by 
$W_\alpha = i\bar{\mc D}^2(e^{-gV}\mc D_\alpha e^{gV})$ where $\mc D_\alpha$ and 
$\bar{\mc D}_{\dot\alpha}$ are supercovariant derivatives. This Lagrangian is 
manifestly $\mc N = 1$ supersymmetric being constructed using the $\mc N = 1$ 
superspace prescription. Moreover, the three chiral superfields enjoy manifest 
$SU(3)$ flavor symmetry, but the full $SU(4)$ $R$-symmetry is obscured.

The general Leigh-Strassler theory is parametrized by complexification of the 
gauge coupling $g$ and modification of the superpotential by the substitution
\begin{align}
\label{LSSUPERPOTENTIALSUB}
g\Tr\left(\Phi_1\Phi_2\Phi_3-\Phi_1\Phi_3\Phi_2\right) \mapsto
\kappa\Tr\left(q\Phi_1\Phi_2\Phi_3-
q^{-1}\Phi_1\Phi_3\Phi_2\right)
+\rho\Tr\left(\Phi_1^3+\Phi_2^3+\Phi_3^3\right)\;,
\end{align}
for complex $\kappa,\rho$ and $q$. The deformations break supersymmetry to 
$\mc N = 1$. In order for the deformation to become exactly marginal and thereby
inherit finiteness at the quantum level, the parameters must be highly
constrained. Leigh and Strassler demonstrated the existence of a three-complex
dimensional surface in the coupling constant space of conformally invariant
theories with $\mc N = 1$ supersymmetry. This surface can be defined as a quite
complicated level set $\gamma(\kappa,\rho,q,g) = 0$. The condition however has
to be calculated perturbatively and unfortunately its form is not known beyond
few loops except for very restrictive choices of the couplings.

\subsection{Star product induced $\beta$-deformations}
The Leigh-Strassler deformed theory generated by the superpotential substitution
\eqref{LSSUPERPOTENTIALSUB} includes as a special case a one-parameter family of
theories known as $\beta$-deformations, which extend to quantum finiteness to
all orders in the planar limit \cite{Mauri:2005pa}. The model follows by 
tightening the assumptions on the parameters and setting $\kappa = g$, 
$\rho = 0$ and $q\bar q = 1$, so that for $\beta_R$ real, the new superpotential 
becomes
\begin{align}
\label{BETASUPERPOTENTIAL}
\mc W_\beta = g\Tr\left(e^{i\pi\beta_R}\Phi_1\Phi_2\Phi_3-
e^{-i\pi\beta_R}\Phi_1\Phi_3\Phi_2\right)\;,
\end{align}
whence the undeformed theory is recovered by sending $e^{i\pi\beta_R}\to 1$. 

It turns out that the $\beta$-deformation can be understood in terms of a
special operation between the superfields, namely the star product, which will 
prove invaluable when we evaluate deformations of scattering amplitudes
\cite{Khoze:2005nd}. In order to define the star product it is necessary to 
discuss the left-over symmetries of the $\beta$-deformed Lagrangian. The 
original $R$-symmetry is broken from $SU(4)_R$ to $U(1)_R$, but it is observed 
that the deformed theory in addition is invariant under a global 
$U(1)_1\times U(1)_2$ flavor symmetry of the three chiral superfields. More 
specifically the symmetries can be expressed as
\begin{align}
U(1)_1:\;\; (\Phi_1,\Phi_2,\Phi_3,V) \mapsto
(\Phi_1,e^{i\alpha_1}\Phi_2,e^{-i\alpha_1}\Phi_3,V)\;, \nonumber \\
\label{U1U1SYMS}
U(1)_2:\;\; (\Phi_1,\Phi_2,\Phi_3,V) \mapsto
(e^{-i\alpha_2}\Phi_1,e^{i\alpha_2}\Phi_2,\Phi_3,V)\;,
\end{align}
the vector superfield being neutral under these transformations. The symmetry 
charges are 
\begin{align}
\label{CHIRALFLAVOUR}
Q^{[1]}\equiv (0,+1,-1,0)\;, \quad\;\;
Q^{[2]}\equiv (-1,+1,0,0)\;.
\end{align}

Suppose that $\Phi_i$ and $\Phi_j$ are chiral superfields with flavor symmetry 
charges $Q^{[1,2]}_i$ and $Q^{[1,2]}_j$ respectively. Then the star product 
operation together with the $\beta$-deformed commutator between these fields is 
defined by
\begin{align}
\label{STARPRODUCTDEFINITION}
\Phi_i\star\Phi_j\equiv
e^{i\pi\beta_R\left(Q_i^{[1]}Q_j^{[2]}-Q_i^{[2]}Q_j^{[1]}\right)}\Phi_i\Phi_j\;, 
\quad [\Phi_i,\Phi_j]_\beta \equiv 
e^{i\pi\beta_{ij}}\Phi_i\Phi_j-e^{-i\pi\beta_{ij}}\Phi_j\Phi_i\;,
\end{align}
where $\beta_{ij}$ is the antisymmetric matrix
\begin{align}
\beta_{ij} = -\beta_{ji}\;, \quad 
\beta_{12} = -\beta_{13} = \beta_{23} \equiv \beta_R\;.
\end{align}
For $\beta_R$ real it follows that the star product is simply the usual product 
adjusted by an overall flavor dependent phase factor. Of important properties of 
the star product we mention associativity, which is a consequence of additivity 
of the flavor charges, and noncommutativity, the latter being obvious from the 
definition \eqref{STARPRODUCTDEFINITION}.

Prior to implementation in the $\mc N = 4$ Lagrangian the star product must 
extend to the antichiral superfields $\bar\Phi_1$, $\bar\Phi_2$ and
$\bar\Phi_3$. They carry opposite charges to the chiral superfields under 
the $U(1)_1\times U(1)_2$ symmetry so that $\Phi_i\bar\Phi_i$ is chargeless 
and $\Phi_i\star\bar\Phi_i = \Phi_i\bar\Phi_i$. We therefore note
\begin{align}
\label{CHIRALFLAVOURBAR}
\bar Q^{[1]}\equiv (0,-1,+1,0) = -Q^{[1]}\;, \quad\;\;
\bar Q^{[2]}\equiv (+1,-1,0,0) = -Q^{[2]}\;,
\end{align}
whereby both antichirality and mixed chirality star products become well defined.

With this operation the $\beta$-deformation can be induced in the $\mc N = 4$ 
Lagrangian by substituting all ordinary products between superfields with star 
products, or equivalently replacing all commutators with $\beta$-deformed 
brackets. In particular, the superpotential can be written 
\begin{align}
\mc W_\beta =
g\Tr\left(\Phi_1\star\Phi_2\star\Phi_3-
\Phi_1\star\Phi_3\star\Phi_2\right) =
g\Tr\left(e^{i\pi\beta_R}\Phi_1\Phi_2\Phi_3-
e^{-i\pi\beta_R}\Phi_1\Phi_3\Phi_2\right)\;,
\end{align}
and hence the $\beta$-deformed Lagrangian takes the form
\begin{align}
\mc L_\beta = &
\int d^2\theta d^2\bar\theta\Tr e^{-gV}\bar\Phi^ie^{gV}\Phi_i \nonumber \\ & {}
+\frac{1}{2g^2}\int d^2\theta\Tr W^\alpha W_\alpha
+\left\{g\int d^2\theta\Tr\Phi_1[\Phi_2,\Phi_3]_\beta+\hc\right\}\;.
\end{align}

\begin{figure}[!h]
\centering
\includegraphics[scale=0.6]{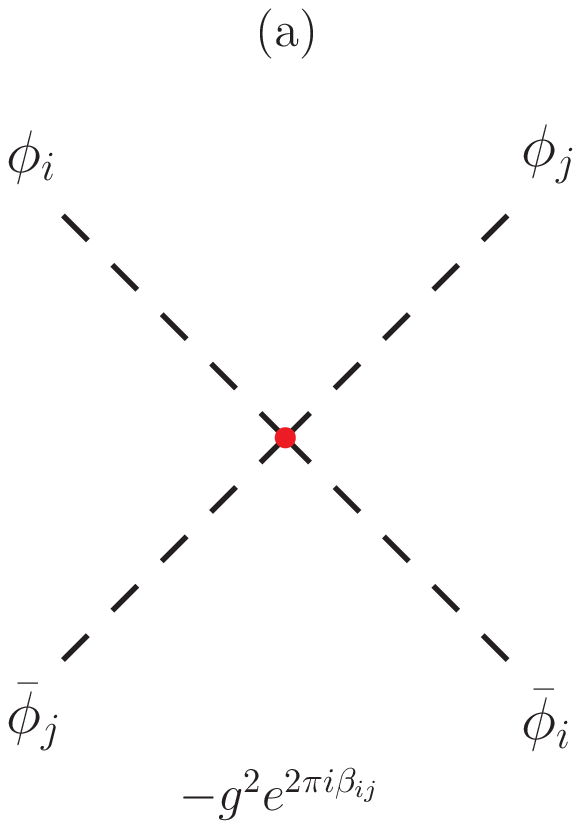} \hspace*{0.5cm}
\includegraphics[scale=0.6]{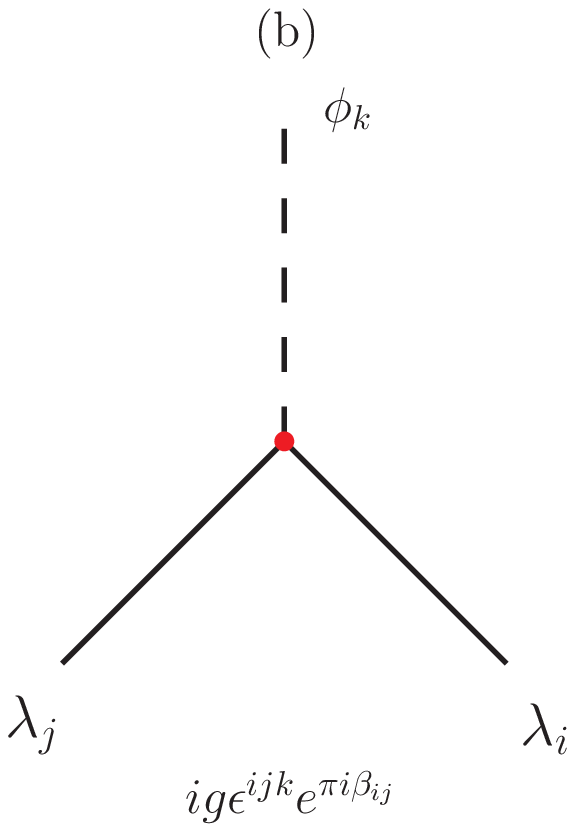} \hspace*{0.5cm}
\includegraphics[scale=0.6]{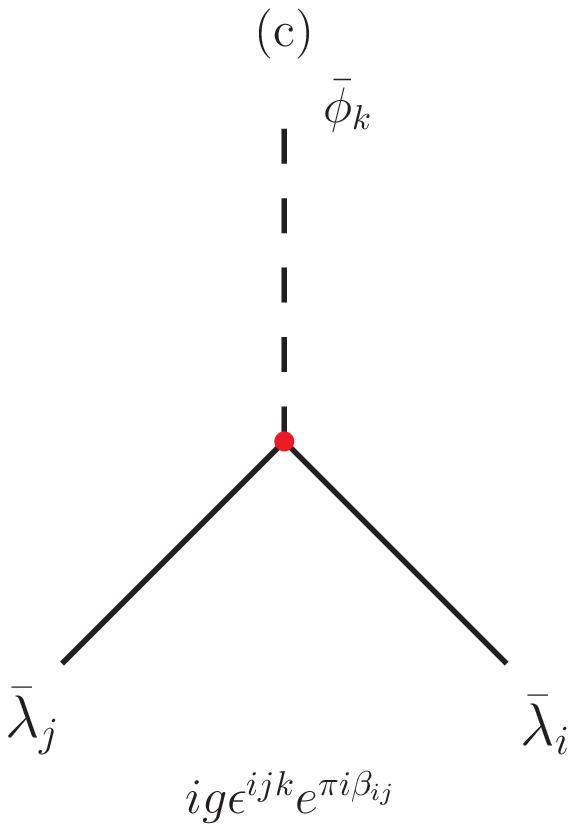}
\caption[$\beta$-dependent Feynman vertices]
{\label{BETADEPENDENTVERTICES}
The $\beta$-deformation adds overall phase factors to the original vertex rules 
of the depicted four-scalar and Yukawa couplings. We denote the three complex 
scalar field components of the chiral superfields $\Phi_i$ by $\phi_i$, while 
$\lambda_i$ label their fermionic superpartners. Notice that the ordering of the 
legs is crucial.}
\end{figure}

By consideration of the component version of the Lagrangian for the $\mc N = 4$
super Yang-Mills theory rather than its superspace representation
\eqref{N=4SYMLAGRANGIAN} it is elementary to derive the modified color-ordered
Feynman rules. We mention that non-consecutive four-scalar interactions and 
Yukawa vertices without particles from the vector multiplet exhaust the 
$\beta$-deformed vertices (see fig.~\ref{BETADEPENDENTVERTICES}), and refer the 
reader to \cite{Khoze:2005nd} for detailed calculations.

\subsection{Phase structure and effective vertices}
Scattering amplitudes in $\mc N = 4$ and $\beta$-deformed super Yang-Mills
theories seem closely related when comparing their Feynman rules. Indeed, only
three specific four-scalar and Yukawa interactions are modified, and this
deformation introduces nothing but prefactors to the vertex rules. Several
comments are important in this connection.

The two theories have the same particle content and it is easy to see that many
amplitudes are actually identical. For a $\beta$-deformed amplitude to agree
with the corresponding $\mc N = 4$ expression, the phases of the
$\beta$-dependent vertices should cancel each other, or such vertices should
simply be absent in the diagram. Whole classes of amplitudes are insensitive to
the phase deformation, the most obvious being all tree-level amplitudes with
external gluons and gluinos exclusively.

In general, amplitudes in the deformed theory are $\beta$-dependent. The 
modifications however turn out to be surprisingly uncomplicated. At first we 
realize that the original and deformed theories are equivalent at tree-level, up 
to a multiplicative prefactor. This conclusion follows from the simplicity of 
the color-ordered Feynman vertices. But it can be shown that the product of 
phase factors from the individual vertices is independent of the internal 
structure of the diagram. Suppose the arbitrary color-ordered tree-level 
amplitude in consideration has any combination of $n$ external fields and label 
them $\Lambda_1,\Lambda_2,\dots,\Lambda_n$. Then the $\beta$-dependence of this 
amplitude is entirely captured by 
$\Tr\left[\Lambda_1\star\Lambda_2\star\cdots\star\Lambda_n\right]$. Actually 
this statement applies not only to tree-level amplitudes, but is guaranteed to 
hold to all orders in perturbation theory in the planar limit. The claim has 
profound consequences for our applications of $\beta$-deformed super Yang-Mills 
theory. A proof based on effective vertices of only external legs was provided
in \cite{Khoze:2005nd}. Shortly we report a similar strategy for supervertices.

We end this section by deriving an expression for the phase factor of any given 
$n$-point amplitude in terms of only the $U(1)_1\times U(1)_2$ charges of the
superfields corresponding to the external legs. Upon invoking the definition of 
the star product \eqref{STARPRODUCTDEFINITION} we deduce the generalization to 
$n$ legs,
\begin{align}
\label{PHASEALLORDERS}
\P^{\beta_R}\left(\Lambda_1,\Lambda_2,\dots,\Lambda_n\right) & 
\equiv \frac{\Tr\left[\Lambda_1\star\Lambda_2\star\cdots\star\Lambda_n\right]}
{\Tr\left[\Lambda_1\Lambda_2\cdots\Lambda_n\right]}
\nonumber \\ & =
\prod_{i=1}^{n-1}\exp\bigg[i\pi\beta_R\sum_{j=i+1}^n\left(
Q^{[1]}_{\Lambda_i}Q^{[2]}_{\Lambda_j}-
Q^{[2]}_{\Lambda_i}Q^{[1]}_{\Lambda_j}\right)\bigg]\;,
\end{align}
using the rather selfexplanatory notation with $Q^1_{\Lambda_i}$ and
$Q^2_{\Lambda_i}$ denoting the symmetry charges of the field $\Lambda_i$.
Finally \eqref{PHASEALLORDERS} may be recast perhaps more conveniently as
\begin{align}
\P^{\beta_R}\left(\Lambda_1,\Lambda_2,\dots,\Lambda_n\right) =
\exp\bigg[i\pi\beta_R\sum_{i<j}\left(
Q^{[1]}_{\Lambda_i}Q^{[2]}_{\Lambda_j}-
Q^{[2]}_{\Lambda_i}Q^{[1]}_{\Lambda_j}\right)\bigg]\;.
\label{PHASEFROMCHARGES}
\end{align}
This result is essential for the rest of the paper.

%%%%%
\section{Generating functions from bilocal phase operators}
Now that we have gained confidence with the basic structure and interactions of
$\beta$-deformed super Yang-Mills theory, it is very natural to attempt to
incorporate all scattering amplitudes sectorwise into generating functions
instead of relying on traditional Feynman calculations. Thereby established 
$\mc N = 4$ superspace applications such as intermediate state sums by Grassmann
integration become compatible with the $\beta$-deformed amplitudes.

We first identify particles in $\mc N = 4$ on-shell superspace with the 
components of the $\mc N = 1$ vector and chiral superfields. Recall that the 
sixteen physical states in the $\mc N = 4$ supermultiplet can be realized as two 
gluons $g_+$ and $g^{abcd}_-$, four fermion pairs $f^a_+$ and $f^{abc}_-$, plus 
six real, self-dual scalars $s^{ab}$, all completely antisymmetric in the 
displayed fundamental $SU(4)$ indices. The gluons of course belong to the vector 
multiplet, while the remaining fermions and scalars of the theory can be chosen 
such that \cite{Ananth:2006ac}
\begin{align}
\left\{
f_+^a,f_-^{abc},s^{i4},{s}^{ij}\right\}\longleftrightarrow
\left\{\lambda^a,\epsilon^{abcd}\bar\lambda_d,
\phi^i,\epsilon^{ijk4}\bar\phi_k\right\}\;,
\end{align}
for $a,b,c = 1,2,3,4$ and $i,j,k = 1,2,3$.

\subsection{MHV generating functions}
We consider the $\MHV$ sector and derive an expression for the $n$-point
tree-level $\MHV$ generating function. With that result at hand the $\gMHV$ 
superamplitude will follow immediately from the Fourier connection. Clearly, 
both expressions should reduce to the original $\mc N = 4$ superamplitudes in 
the limit $\beta_R\to 0$.

The upshot of the preceding section was that the $\mc N = 4$ and 
$\beta$-deformed theories have identical planar sectors up to simple phase 
factors, which for any given amplitude to all orders in perturbation theory are
determined by the configuration of its external legs and their 
$U(1)_1\times U(1)_2$ symmetry charges according to \eqref{PHASEFROMCHARGES}.
The logical solution is therefore to take the $\mc N = 4$ superamplitude and
just attach to each component the appropriate phase factor. 

We remind ourselves that the $\mc N = 4$ $\MHV$ superamplitude reads
\begin{align}
\A^{\mathrm{MHV}}_n = 
i\frac{\prod_{a=1}^4\sum_{i<j}\avg{q_{ia}q_{ja}}}{\prod_{r=1}^n\avg{r(r+1)}}\;,
\end{align}
and therefore the task concentrates on translating the individual Grassmann
signatures to $\beta$-dependent expressions. We circumvent this obstacle by
assigning $U(1)_1\times U(1)_2$ charges to the fermionic coordinates of the 
$\mc N = 4$ on-shell superspace instead of the superfields such that the
on-shell superspace version of the star product imitates the star product 
defined in $\mc N = 1$ superspace. This solution is closely related to the 
light cone superspace star product introduced in \cite{Ananth:2006ac}. We apply
the usual charges $Q^{[1]}$ and $Q^{[2]}$, with opposite signs in the Fourier 
transformed superspace. More precisely $\eta_{ia}$ carries charges $Q^{[1]}_a$ 
and $Q^{[2]}_a$.

The trick is now to construct a pair of differential operators to decode the 
$\eta$-patterns and thereby count symmetry charges. Preferably the 
$\eta$-strings should be eigenstates of these operators, and the symmetry 
charges the corresponding eigenvalues. In order to achieve this we introduce 
the operator
\begin{align}
\oQ^{[1,2]}_i \equiv
\sum_{a=1}^4 Q^{[1,2]}_a\eta_{ia}\partial_{ia}
\end{align}
with the two explicit components
\begin{align}
\oQ^{[1]}_i \equiv
\eta_{i2}\partial_{i2}-
\eta_{i3}\partial_{i3}\;, \quad
\oQ^{[2]}_i \equiv
\eta_{i2}\partial_{i2}-
\eta_{i1}\partial_{i1}
\end{align}
such that $\oQ^{[1]}_i$ and $\oQ^{[2]}_i$ measure the symmetry 
charges $Q_i^{[1]}$ and $Q_i^{[2]}$ respectively for leg $i$, and are by 
construction Grassmann even. For instance the action on the Grassmann 
combination corresponding to the negative helicity fermion of the chiral 
superfield $\Phi_3$ is $\oQ^{[1]}_i(\eta_{i1}\eta_{i2}\eta_{i4}) = 
\eta_{i1}\eta_{i2}\eta_{i4}$ and 
$\oQ^{[2]}_i(\eta_{i1}\eta_{i2}\eta_{i4}) = 0$ in agreement with
\eqref{CHIRALFLAVOURBAR}. It is easy to see that the mechanism works in 
general. The vector multiplet in particular has vanishing charges as it should.

Substituting $\oQ^{[1]}_i$ and $\oQ^{[2]}_i$ for the constant 
symmetry charges in \eqref{PHASEFROMCHARGES} yields the operator version of the 
phase factor
\begin{align}
\label{PHASEFACTOROPERATOR}
\oP^{\beta_R}_n \equiv
\exp\bigg[i\pi\beta_R\sum_{p<q}\left(
\oQ_p^{[1]}\oQ_q^{[2]}-
\oQ_p^{[2]}\oQ_q^{[1]}\right)\bigg]\;,
\end{align}
which enables us to formally write the $\beta$-deformed $\MHV$ superamplitude as
\begin{align}
\A_n^{\beta_R,\,\MHV} & = 
\exp\bigg[i\pi\beta_R\sum_{p<q}\left(
\oQ_p^{[1]}\oQ_q^{[2]}-
\oQ_p^{[2]}\oQ_q^{[1]}\right)\bigg]
\A_n^{\MHV} \nonumber \\ & =
i\prod_{r=1}^n\avg{r(r+1)}^{-1}
\exp\bigg[i\pi\beta_R\sum_{p<q}\left(
\oQ_p^{[1]}\oQ_q^{[2]}-
\oQ_p^{[2]}\oQ_q^{[1]}\right)\bigg]
\prod_{\ell=1}^4\sum_{i<j}\avg{ij}\eta_{i\ell}\eta_{j\ell}\;.
\end{align}
The $\beta$-deformed generating function has by construction the desired 
property $\A_n^{\beta_R,\,\MHV} \to \A_n^{\MHV}$ in the limit 
$\beta_R\to 0$.

In order to streamline the notation we first introduce an alternative version of
the Kronecker delta function defined by
\begin{align}
\delta_{i\{I\}} = \left\{
\begin{array}{ll}
1 & \quad\mathrm{if}\; i\in I\;, \\
0 & \quad\mathrm{otherwise}\;,
\end{array}
\right.
\end{align}
where for our purposes the set $I$ should only contain unique elements. Let us
then turn to the evaluation of the action of the symmetry charge operators on a
generic string of Grassmann variables present in the $\MHV$ superamplitude. The
vector multiplet sector commutes right through the differentiation, while
numerous Kronecker delta functions are produced when hitting the 
$\eta$-variables pertinent to the chiral multiplets. Keeping track of all 
possible combinations we find
\begin{align}
\MoveEqLeft
\left(\oQ_p^{[1]}\oQ_q^{[2]}-
\oQ_p^{[2]}\oQ_q^{[1]}\right)
\prod_{\ell=1}^4\eta_{i_\ell \ell}\eta_{j_\ell \ell} = \nonumber \\
& \left[
(\delta_{p\{i_2,j_2\}}-\delta_{p\{i_3,j_3\}})
(\delta_{q\{i_2,j_2\}}-\delta_{q\{i_1,j_1\}})
-(p\leftrightarrow q)
\right]
\prod_{\ell=1}^4\eta_{i_\ell \ell}\eta_{j_\ell \ell}\;.
\end{align}
It immediately follows that 
$\delta_{p\{i_2,j_2\}}\delta_{q\{i_2,j_2\}}-(p\leftrightarrow q) = 0$. The phase 
factor therefore reduces slightly into
\begin{align}
\label{MHVPHASEFACTOR}
\P^{\beta_R,\,\MHV}_{i_1j_1;i_2j_2;i_3j_3} & \equiv
\exp\bigg(
i\pi\beta_R
\sum_{p<q}
\left[
\delta_{q\{i_1,j_1\}}\left(
\delta_{p\{i_3,j_3\}}-\delta_{p\{i_2,j_2\}}
\right)-\delta_{q\{i_2,j_2\}}\delta_{p\{i_3,j_3\}}
-(p\leftrightarrow q)\right]
\bigg)
\end{align}
such that the $\beta$-deformed $\MHV$ superamplitude becomes
\begin{align}
\label{BETAMHVSUPERAMPLITUDE}
\A_n^{\beta_R,\,\MHV} = {} &
i\prod_{r=1}^n\avg{r(r+1)}^{-1}
\sum_{\{i\}<\{j\}}
\P^{\beta_R,\,\MHV}_{i_1j_1;i_2j_2;i_3j_3} 
\prod_{\ell=1}^4
\avg{q_{i_\ell \ell}q_{j_\ell \ell}}\;,
\end{align}
with the shorthand notation
\begin{align}
\sum_{\{i\}<\{j\}} \equiv \sum_{i_1<j_1}\cdots\sum_{i_4<j_4}\;.
\end{align}

The form of \eqref{BETAMHVSUPERAMPLITUDE} coincides with our expectations. We
see that the three spin factors corresponding to the chiral multiplets are now
correlated through a phase matrix $\P^{\beta,\MHV}_{i_1j_1;i_2j_2;i_3j_3}$ 
while the original fourth $SU(4)$ factor identified with the vector multiplet is 
left unchanged and can be separated out. Hence, $\mc N = 1$ supersymmetry is
manifest. 

The result also reflects that alternatively we could have considered a theory of 
the three chiral superfields with the appropriate phase dependent interactions 
and then have coupled the corresponding superspace structure to the bare 
$\mc N = 1$ $\MHV$ superamplitude addressed in 
\cite{Bern:2009xq,Elvang:2011fx,Sogaard:2011pr}. However, the approach presented 
here more efficiently generalizes to superamplitudes beyond the $\MHV$ sector.

\subsection{Cyclicity and color reflection identities}
Before we continue let us pause for a second and study the phase operator 
defined on on-shell superamplitudes in phase-deformed on-shell superspace. 
Formally, it is given in terms of its Taylor series in the bilocal pure phase 
$\hat\varphi$ such that 
\begin{align}
\oP^{\beta_R} = \exp[i\pi\beta_R\hat\varphi]\;.
\end{align}

The phase generator comes with canonical ordering of the external legs.
However, with the $(n-1)!$ different configurations introduced by color
decomposition in mind, it is necessary to define the operator for an arbitrary
permutation $\sigma$ of $(1,2,\dots,n)$, i.e.
\begin{align}
\oP_\sigma^{\beta_R} =
\exp\bigg[i\pi\beta_R\sum_{p<q}\left(
\oQ_{\sigma(p)}^{[1]}\oQ_{\sigma(q)}^{[2]}-
\oQ_{\sigma(p)}^{[2]}\oQ_{\sigma(q)}^{[1]}\right)\bigg]\;.
\end{align}
With this formulation the phase generator can be brought to depend only on the 
positions of the particles in the cyclic chain and is thus universal.

Amplitudes should have cyclic symmetry. We prove the crucial property of 
cyclicity of phase operator itself. In general, bilocal operators such as the 
phase generator map cyclic functions to non-cyclic functions. However, 
depending on the space upon which the operators act, the non-cyclic remainder 
may be brought to vanish. The most prominent example is the level-one Yangian 
generators.

It is straightforward to see that
\begin{align}
\label{PHASEDIFFCYCLIC}
\sum_{1\leq p<q\leq n}\left(
\oQ_p^{[1]}\oQ_q^{[2]}-
\oQ_p^{[2]}\oQ_q^{[1]}\right)-
\sum_{2\leq p<q\leq 1}\left(
\oQ_p^{[1]}\oQ_q^{[2]}-
\oQ_p^{[2]}\oQ_q^{[1]}\right) =
2\sum_{p=1}^n\bigg(
\oQ_1^{[1]}\oQ_p^{[2]}-
\oQ_1^{[2]}\oQ_p^{[1]}\bigg)\;,
\end{align}
where the two indicated configurations $1,2,\dots,n$ and $2,3,\dots,n,1$ differ
by a cyclic transformation. The on-shell amplitudes are neutral with respect to
the flavor symmetry, which means that
\begin{align}
\sum_{p=1}^n\oQ_p^{[1]}\A_n^{\beta_R} =
\sum_{p=1}^n\oQ_p^{[2]}\A_n^{\beta_R} = 0\;.
\end{align}
Therefore the restriction of the phase factor difference \eqref{PHASEDIFFCYCLIC}
to this space vanishes. Moreover, the remainder term at arbitrary order in the
deformation parameter is bound to annihilate the amplitudes. We thus conclude 
that $\hat\varphi(1,2,\dots,n) = \hat\varphi(2,3,\dots,n,1)$ and
\begin{align}
\A_n^{\beta_R}(1,2,\dots,n) =
\A_n^{\beta_R}(2,3,\dots,n,1)\;.
\end{align}

Let us next consider the phase-deformed color reflection identity. It is easy
to realize that reversal of the order of the external legs inverts the phase,
\begin{align}
\hat\varphi(1,2,\dots,n) = -\hat\varphi(n,n-1,\dots,1)\;.
\end{align}
Remembering that inversion of undeformed amplitudes introduces a factor of
$(-1)^n$ we thus realize that the phase-deformed analogue has to be phase 
dressed. Alternatively, we can compensate for the transformation of the phase 
via the deformation parameter,
\begin{align}
\A_n^{\beta_R}(1,2,\dots,n) = 
(-1)^n\A_n^{(-\beta_R)}(n,n-1,\dots,1)\;.
\end{align}

\subsection{Components of the MHV sector}
We calculate a number of phase factors using \eqref{MHVPHASEFACTOR} to obtain
explicitly some of the components of the $\beta$-deformed MHV generating 
function. Our results agree with the phase structure obtained via usual 
star product defined for $\mc N = 1$ superfields. The relevant phase matrix
indices for the amplitudes written below are $24;14;23$, $13;12;12$, $34;14;13$, 
$34;13;14$, $16;35;56$ and $16;56;35$ respectively.
\begin{align*}
A_4^{\beta_R,\,\MHV}(
1_{s^{24}},2_{s^{13}},3_{s^{34}},4_{s^{12}}) & =
i\frac{\avg{q_{21}q_{41}}\avg{q_{12}q_{42}}\avg{q_{23}q_{33}}\avg{q_{14}q_{34}}}
{\prod_{r=1}^4\avg{r(r+1)}} \\
%%%
A_4^{\beta_R,\,\MHV}(
1^-_{g^{1234}},2^-_{f^{234}},3^+_{f^1},4^+_{g}) & =
i\frac{\avg{q_{11}q_{31}}\avg{q_{12}q_{22}}\avg{q_{13}q_{23}}\avg{q_{14}q_{24}}}
{\prod_{r=1}^4\avg{r(r+1)}} \\
%%%
A_5^{\beta_R,\,\MHV}(
1^-_{f^{234}},2_g^+,3^-_{f^{134}},4_{s^{12}},5^+_{g}) & =
ie^{+i\pi\beta_R}
\frac{\avg{q_{31}q_{41}}\avg{q_{12}q_{42}}\avg{q_{13}q_{33}}\avg{q_{14}q_{34}}}
{\prod_{r=1}^5\avg{r(r+1)}} \\
%%%
A_5^{\beta_R,\,\MHV}(
1^-_{f^{234}},2_g^+,3^-_{f^{124}},4_{s^{13}},5^+_{g}) & =
ie^{-i\pi\beta_R}
\frac{\avg{q_{31}q_{41}}\avg{q_{12}q_{32}}\avg{q_{13}q_{43}}\avg{q_{14}q_{34}}}
{\prod_{r=1}^5\avg{r(r+1)}} \\
%%%
A_6^{\beta_R,\,\MHV}(
1_{s^{14}},2_g^+,3_{s^{24}},4_g^+,5_{s^{23}},6_{s^{13}}) & =
ie^{+2i\pi\beta_R}
\frac{\avg{q_{11}q_{61}}\avg{q_{32}q_{52}}\avg{q_{53}q_{63}}\avg{q_{14}q_{34}}}
{\prod_{r=1}^6\avg{r(r+1)}} \\
%%%
A_6^{\beta_R,\,\MHV}(
1_{s^{14}},2_g^+,3_{s^{34}},4_g^+,5_{s^{23}},6_{s^{12}}) & =
ie^{-2i\pi\beta_R}
\frac{\avg{q_{11}q_{61}}\avg{q_{52}q_{62}}\avg{q_{33}q_{53}}\avg{q_{14}q_{34}}}
{\prod_{r=1}^6\avg{r(r+1)}}
\end{align*}

\subsection{All googly-MHV tree amplitudes}
The $\beta$-deformed $\MHV$ superamplitude allows an almost trivial continuation
to the $\gMHV$ superspace. It just amounts figuring out an expression for the
phase factors. But because the signs of all $U(1)_1\times U(1)_2$ charges just
get flipped, and thus cancel in \eqref{PHASEFROMCHARGES}, the phase factors of 
the two sectors are identical in form,
\begin{align}
\P^{\beta_R,\,\gMHV}_{i_1j_1;i_2j_2;i_3j_3} =
\P^{\beta_R,\,\MHV}_{i_1'j_1';i_2'j_2';i_3'j_3'}\;.
\end{align}
Completely analogous to \eqref{BETAMHVSUPERAMPLITUDE}, implementation of the
phase factor matrix $\P^{\beta_R,\,\gMHV}_{i_1j_1;i_2j_2;i_3j_3}$ in the 
original $\mc N = 4$ $\gMHV$ superamplitude \eqref{BARMHVSUPERAMPLITUDE} 
therefore yields the $\beta$-deformed $\gMHV$ generating function
\begin{align}
\A_n^{\beta_R,\,\gMHV} = {} &
i(-1)^n\prod_{r=1}^n\Avg{r(r+1)}^{-1}
\sum_{\{i\}<\{j\}}
\P^{\beta_R,\,\gMHV}_{i_1j_1;i_2j_2;i_3j_3} 
\prod_{\ell=1}^4
\Avg{\tilde q_{i_\ell}^\ell \tilde q_{j_\ell}^\ell}\;,
\end{align}
or, alternatively using the Grassmann Fourier transform, in holomorphic 
superspace,
\begin{align}
\A_n^{\beta_R,\,\gMHV}(\lambda,\tilde\lambda,\eta) =
i(-1)^n\prod_{r=1}^n\Avg{r(r+1)}^{-1}
\sum_{\{i\}<\{j\}}
\P^{\beta_R,\,\gMHV}_{i_1j_1;i_2j_2;i_3j_3} 
\prod_{\ell=1}^4
\mc E_{n;\ell}(i_\ell,j_\ell)
\Avg{i_\ell j_\ell}\;,
\end{align}
for $\mc E_{n;\ell}$ defined by
\begin{align}
\mc E_{n;\ell}(i,j) \equiv
\frac{1}{(n-2)!}
\sum_{k_1,k_2,\dots,k_{n-2}}
\epsilon^{ijk_1k_2\cdots k_{n-2}}
\eta_{k_1\ell}\eta_{k_2\ell}\cdots\eta_{k_{n-2}\ell}\;.
\end{align}

\section{Vertex expansions, recursion relations and all sectors}
\label{FACTORIZATIONSEC}
Any pattern of Grassmann superspace variables may be mapped to a definite phase
factor, allowing easy extension to non-MHV amplitudes. Consider in full 
generality the $\mc N = 4$ $\NKMHV$ tree-level superamplitude, denoted 
$\A_n^{\NKMHV}$ as usual. In this formal development the precise expression for 
this superamplitude is not important. Upon application of the phase factor 
operator \eqref{PHASEFACTOROPERATOR} to $\A_n^{\NKMHV}$ the $\beta$-deformed 
$\NKMHV$ superamplitude can be reached in the form
\begin{align}
\A_n^{\beta_R,\,\NKMHV} \equiv
\oP^{\beta_R}_n \A_n^{\NKMHV} = 
\exp\bigg[i\pi\beta_R\sum_{p<q}\bigg(
\oQ_p^{[1]}\oQ_q^{[2]}-
\oQ_p^{[2]}\oQ_q^{[1]}\bigg)\bigg]
\A_n^{\NKMHV}\;.
\end{align} 

The $\NKMHV$ superamplitude has Grassmann degree $8+4K$ and each component thus 
carries a Grassmann string with $(2+K)\times 4$ distinct indices. The general 
phase matrix therefore has $(2+K)\times 3$ labels, but the form is completely 
similar to \eqref{MHVPHASEFACTOR}. The sets associated with the Kronecker delta 
functions will just have $2+K$ unique elements each.

\subsection{The CSW superrules}
To be more specific we will establish the MHV vertex expansion of Cachazo, 
Svrcek and Witten (CSW). Our proof is the generating function analogue of 
\cite{Khoze:2005nd}, now formulated in terms of bilocal operators. The 
important point is that neutrality of all vertices implies neutrality of any 
amplitude.

Let us quickly refresh our memory of the CSW rules for constructing the $\NKMHV$
generating tree. The procedure is to draw all tree graphs with $(K+1)$ vertices,
distribute $n$ color-ordered legs, to each of the vertices associate a $\MHV$
superamplitude and finally connect them by a scalar Feynman propagator and for
consistency equate the Grassmann coordinates on both ends of the internal lines
between them. It is now rather elementary to extract all contributions within a
particular topology using Grassmann integration over the $K$ internal lines. The
MHV superrules therefore translate into
\begin{align}
\A_n^{\NKMHV} = i^K\sum_{\mathrm{all}\;\mathrm{graphs}}
\int\bigg[\prod_{i=1}^K\frac{d^4\eta_i}{P^2_i}\bigg]
\A_{(1)}^{\MHV}\A_{(2)}^{\MHV}\cdots
\A_{(I)}^{\MHV}\A_{(K+1)}^{\MHV}
\end{align}
where the discrete sum over all graphs incorporates inequivalent topologies.
Although suppressed here it is important to realize that $P_i$ is an off-shell
momentum. However, in order to have a well-defined product of on-shell trees,
the momenta that enters the $(K+1)$ superamplitudes must be null-projections
constructed from the corresponding off-shell momenta using an arbitrary null
reference vector.

It suffices to consider the $\NMHV$ case to exhaust the general factorization
pattern. By cyclicity of the $\mc N = 4$ superamplitudes we can without loss of
generality arrange their internal legs alternating first and last. Thus the
lines of the two supertrees can be labelled $1,2,\dots,k$ and $k,k+1,\dots,n$
respectively. The sum of the associated phases is
\begin{align}
\hat\varphi_{(1)}+\hat\varphi_{(2)} = 
\sum_{p=1}^{k-1}\sum_{q=p+1}^k
\bigg(\oQ_p^{[1]}\oQ_q^{[2]}-
\oQ_p^{[2]}\oQ_q^{[1]}\bigg)+
\sum_{p=k}^{n-1}\sum_{q=p+1}^n
\bigg(\oQ_p^{[1]}\oQ_q^{[2]}-
\oQ_p^{[2]}\oQ_q^{[1]}\bigg)\;.
\end{align}
We relate this expression to the phase of the full tree, e.g. legs
$1,2,\dots,k-1,k+1,\dots,n$, and observe that
\begin{align}
\hat\varphi_{(1)}+\hat\varphi_{(2)} =
\hat\varphi_{(1+2)}-
\sum_{p=1}^{k-1}\sum_{q=k+1}^n
\bigg(\oQ_p^{[1]}\oQ_q^{[2]}-
\oQ_p^{[2]}\oQ_q^{[1]}\bigg)\;.
\end{align}
By flavor charge conservation at each $\MHV$ supertree the displayed double sum 
annihilates the expanded tree, and it can thus be discarded,
\begin{align}
\sum_{p=1}^{k-1}\sum_{q=k+1}^n
\bigg(\oQ_p^{[1]}\oQ_q^{[2]}-
\oQ_p^{[2]}\oQ_q^{[1]}\bigg)
\A_{(1)}^{\MHV}\A_{(2)}^{\MHV} = 0\;.
\end{align}
The generalization to $K>1$ is straightforward by repetition of the argument. It
follows that the phase factor respects the vertex expansion
\begin{align}
\oP^{\beta_R} = \prod_{\mathrm{supervertices}\; I} \oP_{(I)}^{\beta_R}\;.
\end{align}
We therefore have the $\beta$-deformed MHV vertex expansion
\begin{align}
\A_n^{\beta_R,\;\NKMHV} = i^K\sum_{\mathrm{all}\;\mathrm{graphs}}
\int\bigg[\prod_{i=1}^K\frac{d^4\eta_i}{P^2_i}\bigg]
\A_{(1)}^{\beta_R,\,\MHV}\A_{(2)}^{\beta_R,\,\MHV}\cdots
\A_{(K)}^{\beta_R,\,\MHV}\A_{(K+1)}^{\beta_R,\,\MHV}\;.
\end{align}

Interestingly, this result allows us to circumvent any non-MHV amplitude
deformation calculation by multiplying together simpler MHV phase factors.
What is more, a complete set of four distinct Grassmann variables is also
chargeless, which potentially reduces the extent of nestedness even further.

Moreover, as a passing remark we note that the derivations presented here also 
ensure validity of the phase-dressed super BCFW on-shell recursion relations,
\begin{align}
\A^{\beta_R} = 
\sum_{P_i}\int\frac{d^4\eta_{P_i}}{(2\pi)^4}
\A_L^{\beta_R}(z_{P_i})\frac{i}{P_i^2}\A_R^{\beta_R}(z_{P_i})\;.
\end{align}

\subsection{All NMHV tree amplitudes}
To expose the general pattern and keep tediousness to a minimal extent, we 
resort to the simplest case beyond $\MHV$ level, namely the $\NMHV$ sector, 
calculate the phase factor explicitly and construct the $\beta$-deformed 
generating function. The $\mc N = 4$ $\NMHV$ superamplitude depends on the 
Grassmann object $R_{n;st}$ \eqref{DUALSUPERCONFORMALINV} and is given by the 
compact expression
\begin{align}
\A_n^{\NMHV} = 
i\prod_{r=1}^n\avg{r(r+1)}^{-1}
\prod_{a=1}^4\sum_{i<j}\avg{ij}\eta_{ia}\eta_{ja}
\sum_{1<s<t<n} R_{n;st}\;.
\end{align}
Combination of the two Grassmann sums that appear in $R_{n;st}$ using $s < t$ 
yields
\begin{align}
\A_n^{\NMHV} =
i\prod_{r=1}^n\avg{r(r+1)}^{-1}
\prod_{a=1}^4\sum_{i<j}^n\sum_{k=s}^{n-1}
\avg{ij}\expval{n}{x_{nt}x_{ts}+\theta(t-s)x_{ns}x_{st}}{k}
\eta_{ia}\eta_{ja}\eta_{ka}\;,
\end{align}
with $\theta(x)$ denoting the Heaviside step function with the convention
$\theta(x = 0) = 1$. In order to maintain a hygienic labeling scheme we split
each of the displayed summation indices into four, indicated by a subscript
following the flavor and $R$-symmetry index. The $\NMHV$ phase factor can now
either be derived by again acting with the phase operator, or preferably
inferred from the result for the $\MHV$ superamplitude. We immediately obtain
\begin{align}
\label{NMHVPHASEFACTOR}
\P^{\beta_R,\,\NMHV}_{i_1j_1k_1;i_2j_2k_2;i_3j_3k_3} & \equiv
\exp\bigg(
i\pi\beta_R
\sum_{p<q}
\big[
\delta_{q\{i_1,j_1,k_1\}}\left(
\delta_{p\{i_3,j_3,k_3\}}-\delta_{p\{i_2,j_2,k_2\}}
\right) \nonumber \\ & \;\hspace*{2.7cm}
-\delta_{q\{i_2,j_2,k_2\}}\delta_{p\{i_3,j_3,k_3\}}
-(p\leftrightarrow q)\big]
\bigg)\;.
\end{align}

Before the phase factor is plugged back into the amplitude a slightly 
compressed notation is prepared. We take $R_{n;st}$ and strip off the 
Grassmann delta function to get
\begin{align}
\mc R_{n;st} = \frac{\avg{s(s-1)}\avg{t(t-1)}}
{x_{st}^2\expval{n}{x_{ns}x_{st}}{t}\expval{n}{x_{ns}x_{st}}{t-1}
\expval{n}{x_{nt}x_{ts}}{s}\expval{n}{x_{nt}x_{ts}}{s-1}}\;,
\end{align}
such that $R_{n;st} = \mc R_{n;st}\delta^{(4)}(\Xi_{n;st})$. Furthermore we
introduce the chiral spinor
\begin{align}
\bra{\xi_{n;st}} = \bra{n}x_{nt}x_{ts}+\theta(t-s)\bra{n}x_{ns}x_{st}\;.
\end{align}
Our expression for the $\beta$-deformed $\NMHV$ generating tree is thus
\begin{align}
\A_n^{\beta_R,\,\NMHV} = {} &
i\prod_{r=1}^n\avg{r(r+1)}^{-1}
\sum_{1<s<t<n}
\mc R_{n;st} \nonumber \\ &
\sum_{\{i\}<\{j\}}^n
\sum_{\{k\}=s}^{n-1}
\P^{\beta_R,\,\NMHV}_{i_1j_1k_1;i_2j_2k_2;i_3j_3k_3} 
\prod_{\ell=1}^4
\avg{q_{i_\ell \ell}q_{j_\ell \ell}}\avg{\xi_{n;st}k_\ell}\eta_{k_\ell \ell}\;,
\end{align}
where the sums expand in the obvious ways
\begin{align}
\sum_{\{i\}<\{j\}} \equiv \sum_{i_1<j_1}\cdots\sum_{i_4<j_4}\;, \quad\quad
\sum_{\{k\}=s} \equiv \sum_{k_1=s}\cdots\sum_{k_4=s}\;.
\end{align}

This formula completes our analysis at tree level.

%%%%%
\section{Applications to multi-loop unitarity cuts}
In the following we will look at the general structure of supersymmetric sums in
multi-loop unitarity cuts, which break up phase deformed loop amplitudes into
products of tree amplitudes. Such cuts are optimal to work with instead of using 
lower-loop amplitudes in the construction, because the fully developed tree-level 
formalism is recycled.

We first specialize to plain $\mc N = 4$ theory. Schematically we are interested
in performing the intermediate state sum,
\begin{align}
\sum_{\mathrm{states}}
A_{(1)}^{\mathrm{tree}} A_{(2)}^{\mathrm{tree}}\cdots,
A_{(m)}^{\mathrm{tree}}
\end{align}
for each cut leg. Within the superspace setup this summation is rendered very
elegantly using Grassmann integration over the $\eta$-variables associated with
the internal lines. Let us be a bit more specific and assume that the tree-level 
amplitudes are represented by the superamplitudes $\A_{(m)}^{\mathrm{tree}}$ which 
are connected by $k$ on-shell propagators. All possible internal and external 
particle configurations are then encoded in the supercut
\begin{align}
\mc C^{\mc N = 4} =
\int\bigg[\prod_{i=1}^k d^4\eta_i\bigg]
\A_{(1)}^{\mathrm{tree}}\A_{(2)}^{\mathrm{tree}}\cdots
\A_{(m)}^{\mathrm{tree}}\;.
\end{align}

Without loss of generality the $m$ superamplitudes can be assumed to be either
of $\MHV$ or $\gMHV$ type. This is of course trivially justified if all
tree-level amplitudes in the supercut have at most five legs. In more
complicated situations where this is not the case, the $\MHV$ vertex expansion
applies to reduce non-MHV parts to products of $\MHV$ superamplitudes with
additional propagators. Hence, we only have to consider supercuts of the form
\begin{align}
\mc C^{\mc N = 4} =
\int\bigg[\prod_{i=1}^k d^4\eta_i\bigg]
\A_{(1)}^\MHV\cdots\A_{(m')}^\MHV
\hat{\A}_{(m'+1)}^{\gMHV}\cdots \hat{\A}_{(n)}^{\gMHV}\;,
\end{align}
with $m'$ and $n-m'$ $\MHV$ and $\gMHV$ supertrees respectively. Here, $\gMHV$
superamplitudes have been Fourier transformed from tilded superspace to the
$\eta$-coordinates.

Let us now switch on the deformation. All ingredients are at hand. We exploit
that we can attach a subphase to each supertree $I$ and derive the deformed
supercut
\begin{align}
\mc C^{\beta_R} & =
\int\bigg[\prod_{i=1}^k d^4\eta_i\bigg]
\bigg(\prod_I\oP_{(I)}^{\beta_R}\bigg)
\A_{(1)}^\MHV\cdots\A_{(m')}^\MHV
\A_{(m'+1)}^{\gMHV}\cdots
\A_{(n)}^{\gMHV}\nonumber \\ & = 
\int\bigg[\prod_{i=1}^k d^4\eta_i\bigg]
\A_{(1)}^{\beta_R,\;\MHV}\cdots\A_{(m')}^{\beta_R,\;\MHV}
\A_{(m'+1)}^{\beta_R,\;\gMHV}\cdots
\A_{(n)}^{\beta_R,\;\gMHV}\;.
\end{align}
Of course, for planar graphs the phases combine and reproduce the tree-level
result determined by the external legs. But the deformed supercut provides
rich information about non-planar diagrams which will differ substantially
from the large-$N_c$ limit. Beautiful algebraic and graphical methods for
evaluating such supersums were reported in \cite{Bern:2009xq}. These techniques
rely merely on the Grassmann structure of the amplitudes and are hence directly 
compatible with our results. The reader is encouraged to also consult 
\cite{Jin:2012mk}.

%%%%%
\section{A phase representation of the superconformal algebra}
The remarkable properties of amplitudes in conformal deformations of $\mc N = 4$
super Yang-Mills theory with minimal or no supersymmetry suggest that neither the
ordinary or dual representations of the superconformal algebras are really 
natural frameworks for discussing their symmetries. In the following we therefore
propose a novel phase representation of the $psu(2,2|4)$ algebra.\footnote{This 
section originates from enlightening discussions with Florian Loebbert, whom it 
is a pleasure to thank accordingly.} We draw attention to
\cite{Drummond:2009fd,Bargheer:2009qu,Bargheer:2011mm} for thorough treatments of
superconformal and Yangian symmetry.

Let $J$ be any standard $\mc N = 4$ superconformal symmetry generator, i.e. take
\begin{align}
\mc J\in\{p^{\alpha\dot\alpha},q^{\alpha a},\bar{q}^{\dot\alpha}_A,
m_{\alpha\beta},\bar{m}_{\dot\alpha\dot\beta},r^a_{\;\;b},d,s^\alpha_a,
\bar{s}^{\dot\alpha a},k_{\alpha\dot\alpha}\}\;.
\end{align}
We can very intuitively apply a similarity transformation to obtain a
representation that manifestly annihilates the phase-deformed superamplitudes.
Indeed, we can formally remove the phase, apply the ordinary symmetry generator
and then reinsert the deformation. On the space of amplitudes the phase
generator has a perfectly well-defined and very simple inverse given by
\begin{align}
(\oP^{\beta_R})^{-1} = 
\exp\bigg[-i\pi\beta_R\sum_{p<q}\left(
\oQ_p^{[1]}\oQ_q^{[2]}-
\oQ_p^{[2]}\oQ_q^{[1]}\right)\bigg]\;.
\end{align}
Our phase representation is therefore
\begin{align}
\mc J^{\beta_R} = \P^{\beta_R} \mc J (\P^{\beta_R})^{-1}\;.
\end{align}

The transformed operators trivially satisfy the correct commutator and 
anti-commutator relations of the $\mc N = 4$ superconformal algebra by 
construction, irrespective of $[\mc J,\P]$ being nonzero. Moreover, if $\mc J$
is a symmetry of $\A_n$, that is $\mc J\A_n = 0$, then
\begin{align}\
\mc J^{\beta_R}\A_n^{\beta_R} = 0\;.
\end{align}
In words, the deformed generating trees are manifestly annihilated by all
$psu(2,2|4)$ generators in this phase representation. What is more, we can also
apply the transformation to level one generators and hence lift the $psu(2,2|4)$ 
symmetry algebra to a Yangian realized on the deformed amplitudes. Our
discussion thus suggests that all intrinsic properties of planar $\mc N = 4$ 
super Yang-Mills theory are preserved by the phase deformation.

%%%%%
\section{Amplitudes in the $\gamma$-deformation}
We emphasize that our formalism extends almost effortlessly to the
$\gamma$-deformation, of which the $\beta$-deformation is actually a more
frequently studied special case. It is also conformally invariant at the quantum
level in the planar approximation and can be defined analogously
\cite{Oz:2007qr,Ananth:2007px}.

Indeed, the $\gamma$-deformation is generated by promoting ordinary products in 
the $\mc N = 4$ Lagrangian to star products adjusted with phases which now break 
the $SU(4)_R$ $R$-symmetry to its Cartan subgroup, which is a 
$U(1)_1\times U(1)_2\times U(1)_3$ flavor symmetry of the resulting theory.
Customarily the star product between superfields $\Lambda$ and $\Lambda'$ is
\begin{align}
f\star g =
\exp\bigg(i\pi\gamma_i\epsilon^{ijk}q^f_jq^g_k\bigg)
\end{align}
for some basis $q_1,q_2,q_3$. More practically to us though, the phases can
equivalently be parametrized by four-component charges $\mc U^{[1]}_a$ and 
$\mc U^{[2]}_a$ subject only to tracelessness conditions assuming that all 
$\mc N = 1$ multiplets are charged under the flavor symmetry. The number of 
independent parameters is thus three since the charges enter the star product 
antisymmetrically. On the other hand, if the vector multiplet is neutral, 
$\mc U^{[1]}_4 = \mc U^{[2]}_4 = 0$, we recover the one-parameter 
$\beta$-deformation.

Conservation of charge under each symmetry again implies absense of phase 
contributions from internal structure to all orders in planar perturbation 
theory. It is now trivial to write down the appropriate phase generator
\begin{align}
\oP^\gamma \equiv
\exp\bigg[i\pi\beta_R\sum_{p<q}\bigg(
\hat{\mc U}_p^{[1]}\hat{\mc U}_q^{[2]}-
\hat{\mc U}_p^{[2]}\hat{\mc U}_q^{[1]}\bigg)\bigg]
\end{align}
with charge counting operators defined in on-shell superspace in the usual way. 
The extracted parameter $\beta_R$ is a reminiscence of the $\beta$-deformation
and should just be considered a fixed common constant of proportionality. We can 
now easily derive all tree-level amplitudes, super vertex expansions, multi-loop 
unitarity cuts and so forth. In other words, everything we have said about the 
$\beta$-deformation is compatible with the $\gamma$-deformation.

%%%%%
\section{Summary and outlook}
In this paper we have investigated the perturbative regime of $\beta$-deformed
super Yang-Mills theory using on-shell methods. We have explicitly written all 
MHV and NMHV tree-level scattering amplitudes in terms of new generating 
functions and proven generalization to arbitrary particle and helicity 
configurations via the MHV vertex expansion. Our results have been obtained by 
implementation of a phase matrix in the $\mc N = 4$ superamplitudes, derived from 
their Grassmann structure using a sector independent operator. Several component 
amplitudes were given as examples. 

All generating trees are manifestly $\mc N = 1$ supersymmetric and reduce to the 
usual maximally supersymmetric expressions when the deformation is removed. 
However, we transformed the $\mc N = 4$ superconformal generators in on-shell 
superspace and uncovered a phase dependent representation that annihilates the 
deformed amplitudes. In this implementation, all symmetries exhibited by the 
$\mc N = 4$ amplitudes survive the deformation.

We finally set the stage for automated computation of intermediate state sums in 
connection with multi-loop unitarity cuts of non-planar amplitudes in both the 
$\beta$- and $\gamma$-deformation. Applications of generating functions in this
direction seem especially promising for providing further novel insight.

%%%%%
\begin{acknowledgments}
The author is grateful to Florian Loebbert, Poul Henrik Damgaard and Emil 
Bjerrum-Bohr for many helpful discussions. 
\end{acknowledgments}

%%%%%

\end{document}